\RequirePackage{lineno}
\documentclass[aps,prl,twocolumn,superscriptaddress,showpacs,preprintnumbers,amsmath,amssymb]{revtex4-2}
\usepackage[dvipsnames]{xcolor}
\usepackage[normalem]{ulem}
\usepackage{comment}
\usepackage{overpic}
\usepackage{multirow}
\usepackage{booktabs} 

\graphicspath{{./figures/}}
\usepackage[colorlinks,linkcolor=red,anchorcolor=blue,citecolor=red,breaklinks=true]{hyperref}
\usepackage{cleveref}
\newsavebox{\tablebox}
\usepackage{orcidlink}

\input{alias.tex}
\def\LcTopKs {{\Lambda_c^+\to{}p\KS}}
\def\LcTopKsKs {{\Lambda_c^+\to{}p\KS\KS}}
\def\LcTopKsEta {{\Lambda_c^+\to{}p\KS\eta}}
\def\sPlot {{}_{s}\mathcal{P}lot} 

\begin{document}

\title{\boldmath Measurement of branching fractions of $\Lambda_c^+\to{}pK_S^0K_S^0$ and $\Lambda_c^+\to{}pK_S^0\eta$ at Belle}

\noaffiliation
  \author{L.~K.~Li\,\orcidlink{0000-0002-7366-1307}} 
  \author{K.~Kinoshita\,\orcidlink{0000-0001-7175-4182}} 

  \author{I.~Adachi\,\orcidlink{0000-0003-2287-0173}} 
  \author{J.~K.~Ahn\,\orcidlink{0000-0002-5795-2243}} 
  \author{H.~Aihara\,\orcidlink{0000-0002-1907-5964}} 
  \author{S.~Al~Said\,\orcidlink{0000-0002-4895-3869}} 
  \author{D.~M.~Asner\,\orcidlink{0000-0002-1586-5790}} 
  \author{T.~Aushev\,\orcidlink{0000-0002-6347-7055}} 
  \author{R.~Ayad\,\orcidlink{0000-0003-3466-9290}} 
  \author{V.~Babu\,\orcidlink{0000-0003-0419-6912}} 
% ...
\collaboration{The Belle Collaboration}


\begin{abstract}
We present a study of a singly Cabibbo-suppressed decay $\LcTopKsKs$ and a Cabibbo-favored decay $\LcTopKsEta$
based on $980~\invfb$ of data collected by the Belle detector, operating at the KEKB energy-asymmetric $\epem$ collider.
We measure their branching fractions relative to $\LcTopKs$: 
$\mathcal{B}(\LcTopKsKs)/\mathcal{B}(\LcTopKs)=(1.48\pm0.08\pm0.04)\times10^{-2}$ and $\mathcal{B}(\LcTopKsEta)/\mathcal{B}(\LcTopKs)=(2.73\pm0.06\pm0.13)\times10^{-1}$. 
Combining with the world average $\mathcal{B}(\LcTopKs)$, we have the absolute branching fractions,
$\mathcal{B}(\LcTopKsKs) = (2.35\pm 0.12\pm0.07\pm 0.12)\times 10^{-4}$ and 
$\mathcal{B}(\LcTopKsEta) = (4.35\pm 0.10\pm 0.20 \pm 0.22 )\times 10^{-3}$. 
The first and second uncertainties are statistical and systematic, respectively, while the third ones arise from the uncertainty on $\mathcal{B}(\LcTopKs)$.
The mode $\LcTopKsKs$ is observed for the first time and has a statistical significance of $>\!10\sigma$. The branching fraction of $\LcTopKsEta$ has been measured with a threefold improvement in precision over previous results and
is found to be consistent with the world average. 

\end{abstract}


\maketitle

\section{Introduction}
The weak decays of charmed baryons provide an excellent platform for understanding Quantum Chromodynamics with transitions involving the charm quark. 
The decay amplitudes consist of factorizable and non-factorizable contributions. 
The latter may play a non-trivial or essential role and are approached in various ways, including 
the pole model~\cite{Xu:1992vc,Cheng:1993gf}, 
the covariant confined quark model~\cite{Korner:1992wi,Ivanov:1997ra}, current algebra~\cite{Sharma:1998rd,Zou:2019kzq,Cheng:2018hwl} and SU(3)${}_{\rm F}$ symmetry~\cite{Lu:2016ogy,Geng:2019xbo,Geng:2019awr}. 
To date, there is no established phenomenological model that consistently describes baryon decays.
Precise measurements of branching fractions of charmed baryon weak decays are useful for studying the dynamics of charmed baryons and testing the predictions of theoretical models. 
In addition, the singly Cabibbo-suppressed~(SCS) charm decays are essential probes 
of $\CP$ violation in the charm sector~\cite{Brod:2011re,Cheng:2012wr,Li:2012cfa}
and new physics beyond the standard model~\cite{Grossman:2006jg,Grossman:2012eb,Altmannshofer:2012ur}. 

Experimentally, the investigation of charmed baryons is 
more challenging than that of charmed mesons, mainly due to lower production rates. 
For the lightest state, $\Lcp$, hadronic modes have been studied at several experiments, 
but some have yet to be observed or are measured with low precision~\cite{bib:PDG2022}.
For the Cabibbo-favored~(CF) channel $\LcTopKsEta$~\footnote{Throughout this paper charge-conjugate modes are implied.}, the world average branching fraction, $\BR(\LcTopKsEta)\!=\!{(4.15\pm0.90)\times10^{-3}}$~\cite{bib:PDG2022}, still has a large uncertainty ($22\%$). 
The SCS mode $\LcTopKsKs$, for which the predicted branching fraction is $\BR(\LcTopKsKs)\!=\!{(1.9\pm 0.4)\times10^{-3}}$ based on SU(3)${}_{\rm F}$ symmetry~\cite{Cen:2019ims}, has not previously been observed.

In this paper, we present a precise measurement of $\BR(\LcTopKsKs)$ and $\BR(\LcTopKsEta)$ based on the full Belle data set.
For both of these three-body decays, the Dalitz plot is of interest for the study intermediate resonances.
Understanding the nature of $N^*(1535)$ is very challenging and important for hadronic physics.
The mass of $N^*(1535)$, with spin parity $J^P\!=\!1/2^-$, is larger than that of the radial excitation $N^{*}(1440)$, in opposition to predictions of classical constituent quark models~\cite{Capstick:2000qj}. 
The $N^{*}(1535)$ also couples strongly to channels with strangeness, such as $\eta{}N$ and $K\Lambda$, which is difficult to explain within the naive constituent quark models~\cite{Xie:2017erh,Pavao:2018wdf}. 
The inclusion of five-quark components gives a natural explanation for these properties~\cite{Zou:2010tc}.
The $\LcTopKsEta$ decay, in which the final-state $p\eta$ is in a pure isospin $I\!=\!1/2$ state, is an ideal process for studying the $N^*(1535)$ resonance, as $N^*(1535)$ has a large branching ratio to $p\eta$, in $S$-wave. 
Other intermediate resonances of interest are the light scalars $a_0(980)$ and $f_0(980)$, which both couple to $K\overline{K}$ in $\LcTopKsKs$. They contribute to the SCS $\Lcp$ decays $\Lcp\to{}pK\overline{K}$ and $\Lcp\to{}p\pi\pi$, as predicted in Ref.~\cite{Wang:2020pem}, and likely contribute to $\LcTopKsKs$, based on isospin symmetry.
The nature of $f_0(980)$ and $a_0(980)$ remains poorly understood and continues to be controversial~\cite{Guo:2017jvc,Achasov:2020aun,Wang:2022vga}.
They are often interpreted as compact tetraquark states~\cite{Alford:2000mm,Maiani:2004uc,tHooft:2008rus} or $K\overline{K}$ bound states~\cite{Weinstein:1990gu,Baru:2003qq}. 
Therefore, we reconstruct the Dalitz plots of $\LcTopKsKs$ and $\LcTopKsEta$ decays to check such interesting intermediate resonances.

\section{Detector and data set}
This analysis uses the full dataset recorded by the Belle detector~\cite{belle_detector} 
operating at the KEKB energy-asymmetric $e^+e^-$ collider~\cite{KEKB}. 
This data sample corresponds to a total integrated luminosity of $980~\invfb$ collected at or near the $\Upsilon(nS)$ ($n\!=\!1,\,2,\,3,\,4,\,5$) resonances. 
The Belle detector is a large-solid-angle magnetic spectrometer consisting of a silicon vertex detector, a central drift chamber~(CDC), an array of aerogel threshold Cherenkov counters~(ACC), a barrel-like arrangement of time-of-flight scintillation counters~(TOF), and an electromagnetic calorimeter~(ECL) consisting of CsI(Tl) crystals. These components are all located inside a superconducting solenoid coil that provides a 1.5~T magnetic field. The iron flux-return of the magnet is instrumented to detect $K^0_L$ mesons and to identify muons. The detector is described in detail elsewhere~\cite{belle_detector}.

Monte Carlo (MC) simulated events are generated with {\sc{evtgen}}~\cite{Lange:2001uf} and {\sc{pythia}}~\cite{Sjostrand:2000wi}, 
and are subsequently processed through the full detector simulation based on {\sc{geant3}}~\cite{Brun:1987ma}. 
Final-state radiation from charged particles is included at the event generation stage using {\sc{photos}}~\cite{Barberio:1993qi}.
``Generic" MC samples include $B\overline{B}$ events and continuum processes $e^+e^-\to{}q\bar{q}$ ($q\!=\!u,\,d,\,s,\,c)$ corresponding to an integrated luminosity three times that of the data. 
Samples of MC events of $\Lcp$ signal decay modes are produced in the $e^+e^-\to{}c\cbar$ process, decayed uniformly in three-body phase space, and used to study the efficiency.

\section{Event selection}
We reconstruct the two signal modes $\LcTopKsKs$ and $\LcTopKsEta$ and their reference mode $\LcTopKs$.
The event selections are optimized based on a figure of merit~(FOM), defined as ${\rm FOM}\!=\!\eff_{\scriptscriptstyle{S}}/\sqrt{N_B}$ for $\LcTopKsKs$ due to its branching fraction having not yet been measured, 
and ${\rm FOM}\!=\!N_S/\sqrt{N_S+N_B}$ for $\LcTopKsEta$ assuming its current world average branching fraction~\cite{bib:PDG2022}. 
Here $\eff_{\scriptscriptstyle{S}}$ is the selection efficiency of signal, $N_S$ and $N_B$ are the expected yields of signal and background, respectively, based on
numbers of candidates in the $M(\Lcp)$ signal regions, where $M(\Lcp)$ is the invariant mass of reconstructed $\Lcp$ candidates. 
These signal regions are defined to be within 10, 22, and 18 MeV/$c^2$ of the nominal $\Lcp$ mass~\cite{bib:PDG2022} for the $\LcTopKsKs$, $\LcTopKsEta$, and $\LcTopKs$ channels, respectively; each signal band includes $\approx$98\% of the signal.
For the expected background, $N_B$, the number found in MC is multiplied by the data/MC yield ratio in the $M(\Lcp)$ sideband region ($30<|M(\Lcp)-m_{\Lcp}|\!<\!{50~{\rm MeV}/c^2}$), where $m_{\Lcp}$ is the nominal $\Lcp$ mass~\cite{bib:PDG2022}.

The particle identification (PID) likelihood for a given particle hypothesis, $\mathcal{L}_i$ ($i\!=\!\pi,\,K,\,p$), is calculated from 
the Cherenkov photon yield in the ACC, energy-loss measurements in the CDC, and time-of-flight information from the TOF~\cite{Nakano:2002jw}. 
Charged tracks satisfying $\mathcal{R}(p|K)\!=\!\mathcal{L}_p/(\mathcal{L}_p+\mathcal{L}_K)\!>\!0.6$ and $\mathcal{R}(p|\pi)\!=\!\mathcal{L}_p/(\mathcal{L}_p+\mathcal{L}_{\pi})\!>\!0.6$, are identified as protons. These PID requirements have signal efficiencies of 94\% for $\LcTopKsKs$ and 97\% for $\LcTopKsEta$. 

For proton candidates, the point on the track nearest to the axis defined by the positron beam and in the direction opposite to it (``$z$-axis'') is required to be within  3.0~cm of the interaction point in the $z$-direction and within 1.0~cm on the transverse ($x$-$y$) plane.
This requirement rejects tracks not originating at the interaction point (IP) and introduces a negligible signal efficiency loss ($<0.01\%$).

Candidate $\KS$'s are reconstructed from pairs of oppositely-charged tracks, treated as pions, using an artificial neural network (NN)~\cite{Feindt:2006pm}. 
The NN utilizes the following 13 input variables: 
the $\KS$ momentum in the laboratory frame; 
the separation in $z$ between the two $\pi^{\pm}$ tracks at their intersection in the $x$-$y$ plane; 
for each track, the nearest distance to the IP in the $x$-$y$ plane; 
the $\KS$ flight length in the $x$-$y$ plane; 
the angle between the $\KS$ momentum and the vector joining the IP to the $\KS$ decay vertex; 
in the $\KS$ rest frame, the angle between the $\pi^+$ momentum and the laboratory-frame boost direction; 
and, for each $\pi^{\pm}$ track, the number of CDC hits in both stereo and axial views, and the presence or absence of SVD hits. 
Detailed information is provided elsewhere~\cite{nisKsFinderAtBelle}.
The invariant mass of the reconstructed $\KS\to\pip\pim$ candidate is required to lie within $10~{\rm MeV}/c^2$ of the nominal $\KS$ mass~\cite{bib:PDG2022}; this includes 99.9\% of the $\KS$ signal. 
The two pion tracks from each $\KS$ candidate are refitted to originate from a common vertex and constrained to have invariant mass equal to the nominal $\KS$ mass~\cite{bib:PDG2022}. 
The corresponding fit quality $\chi_{\rm mv}^2(\KS)$ is required to be smaller than 100. 
The selected $\KS$ sample has a purity of greater than 98\%.

Photon candidates are identified as energy clusters in the ECL that are not associated with any charged track. 
The ratio of the energy deposited in the 3$\times$3 array of crystals centered on the crystal with the highest energy, to the energy deposited in the corresponding 5$\times$5 array of crystals, is required to be greater than 0.8. 
The photon energy is required to be greater than 50 MeV in the barrel region (covering the polar angle $32^{\circ}\!<\!\theta\!<\!129^{\circ}$), and greater than 100 MeV in the endcap region ($12^{\circ}\!<\!\theta\!<\!31^{\circ}$ or $132^{\circ}\!<\!\theta\!<\!157^{\circ}$). 

Candidate $\eta\to\gamma\gamma$ decays are reconstructed from photon pairs having an invariant mass satisfying $500~{\rm MeV}/c^2\!<\!M(\gamma\gamma)\!<\!580~{\rm MeV}/c^2$ ($3\sigma$ in $M_\eta(\gamma\gamma)$ resolution). 
The invariant mass of each $\eta$ candidate is constrained to the nominal $\eta$ mass~\cite{bib:PDG2022} at the $\Lcp$ decay vertex (described below). 
The fit quality of this mass constraint is required to satisfy $\chi_m^2(\eta)\!<\!8$, and the resulting $\eta$ momentum in the laboratory frame is required to be greater than 0.4 GeV/$c$. 
To further suppress the background, $\eta$ candidates are vetoed if either of daughters can be paired with another photon such that the $\gamma\gamma$ pair has an invariant mass within $2.5\sigma$ of the nominal $\piz$ mass ($\sigma\!=\!5$ MeV/$c^2$). This $\piz$-veto results in a signal loss of 28\% and removes 72\% of background.

The $\Lcp$ candidates are assembled by forming combinations of the final-state particles for each mode. 
The $p$ and $\KS$ are required to originate from a common vertex (denoted the $\Lcp$ decay vertex and the $\KS$ production vertex) with a fit quality $\chi_{\rm vtx}^2\!<\!24$.
To reduce combinatorial background, the scaled momentum of the $\Lcp$ candidate, defined as $x_p\!=\!p^{*}c/\sqrt{s/4-M^2(\Lcp)\cdot c^4}$, is required to be greater than 0.48, 
where $s$ is the square of the center-of-mass energy and $p^*$ is the momentum of reconstructed $\Lcp$ candidates in the $e^+e^-$ center-of-mass frame.

For the SCS decay $\LcTopKsKs$, a non-$\KS$ peaking background from the CF decay $\Lcp\to p\KS\pip\pim$ exists, even though it is suppressed by the vertex fit and $\KS$ selection.
The $\KS$ decay length $L$ is determined by the projection of the vector joining the $\KS$ production and decay vertices onto the $\KS$ momentum direction, 
and its corresponding uncertainty $\sigma_L$ is calculated by propagating uncertainties in the vertices and the $\KS$ momentum, including their correlations. 
To suppress the non-$\KS$ peaking CF background, we require the significance of the $\KS$ decay length $L/\sigma_L(\KS)\!>\!10$ for the slower of the two $\KS$'s in $\LcTopKsKs$.
This requirement reduces the signal efficiency by 3\%, and rejects 80\% of non-$\KS$ peaking background. 
The remaining non-$\KS$ peaking background is ignored in the $M(\Lcp)$ fits because it has a tiny ratio 0.4\% to signal based on the MC studies with the branching fraction
$(1.6\pm0.12)\%$~\cite{bib:PDG2022}, but considered in the systematic uncertainty.

After applying all selection criteria to the data, we find 1.03, 1.06, and 1.01 candidates per event for $\LcTopKsKs$, $\LcTopKsEta$, and $\LcTopKs$, respectively, in candidates selected from the entire $M(\Lcp)$ fit region ($|M(\Lcp)-m_{\Lcp}|\!<\!0.05~{\rm GeV}/c^2$). 
Correspondingly, about 3.1\%, 5.7\% and 1.2\% of events have multiple signal candidates, which do not introduce any peaking background. 
We retain all candidates for this branching fraction measurement.

\section{Yield extraction}
The signal yield is extracted by an unbinned extended maximum likelihood fit to the $M(\Lcp)$ distribution. 
The signal probability density function (PDF) is a sum of three symmetric Gaussian functions for the $\LcTopKs\KS$ mode, a sum of one symmetric Gaussian and two asymmetric Gaussians for the $\LcTopKsEta$ mode, and a sum of one symmetric Gaussian and three asymmetric Gaussians for the $\LcTopKs$ mode. 
The Gaussian functions share a common mean parameter but have different width parameters. 
The fit is first performed on truth-matched signal MC events.

\begin{figure*}[!htbp]
  \begin{centering}%
  \begin{overpic}[width=0.333\textwidth]{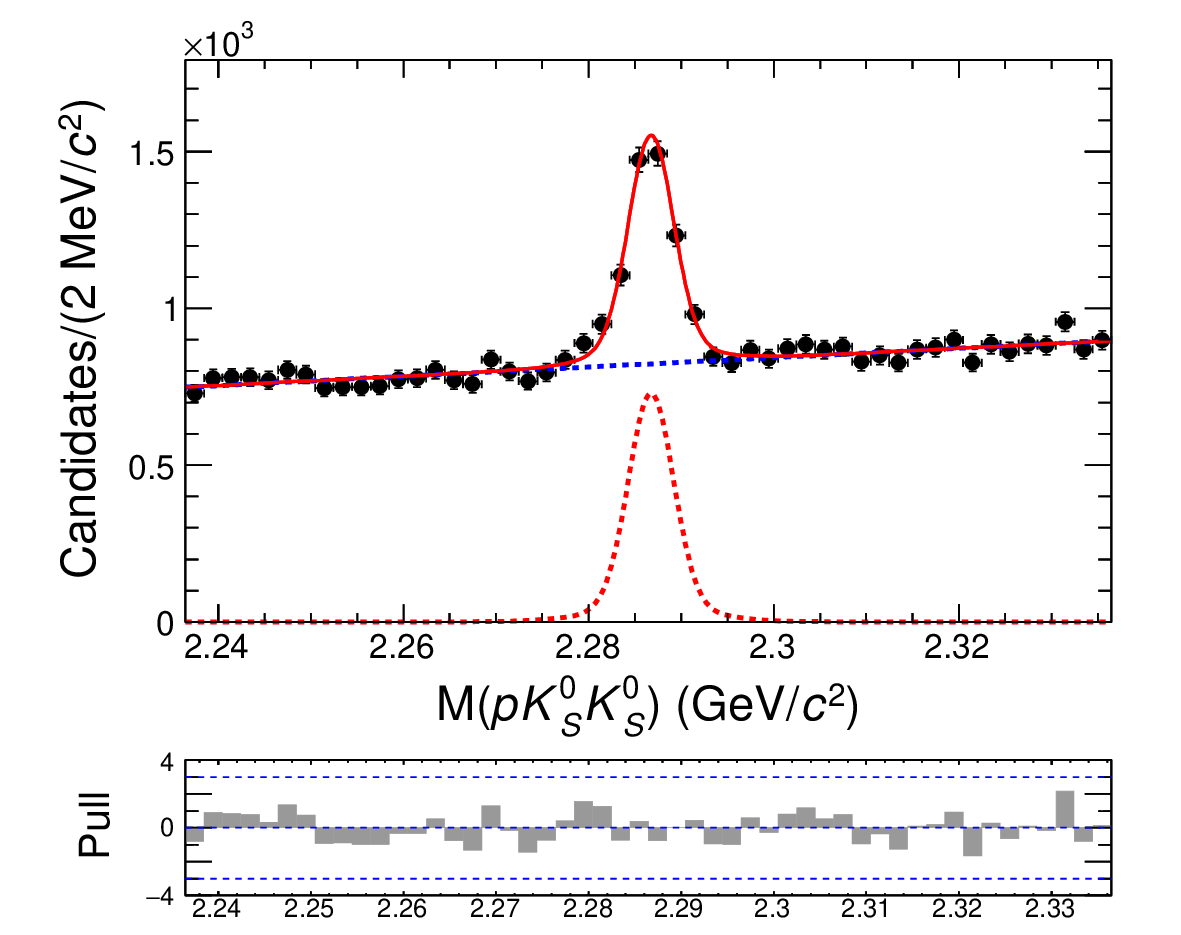}%
  \put(78,65){\large (a)}%
  \end{overpic}%
  \begin{overpic}[width=0.333\textwidth]{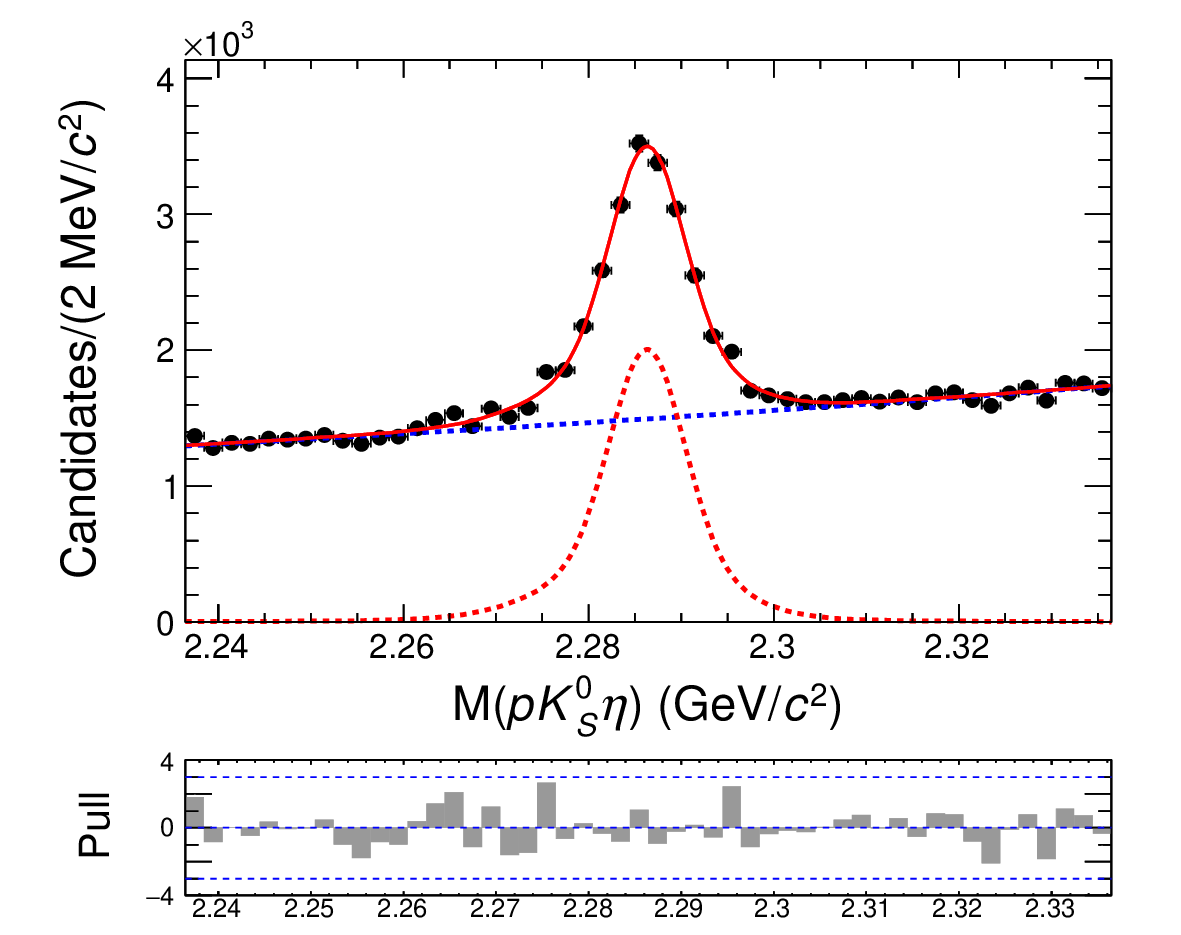}%
  \put(78,65){\large (b)}%
  \end{overpic}%
  \begin{overpic}[width=0.333\textwidth]{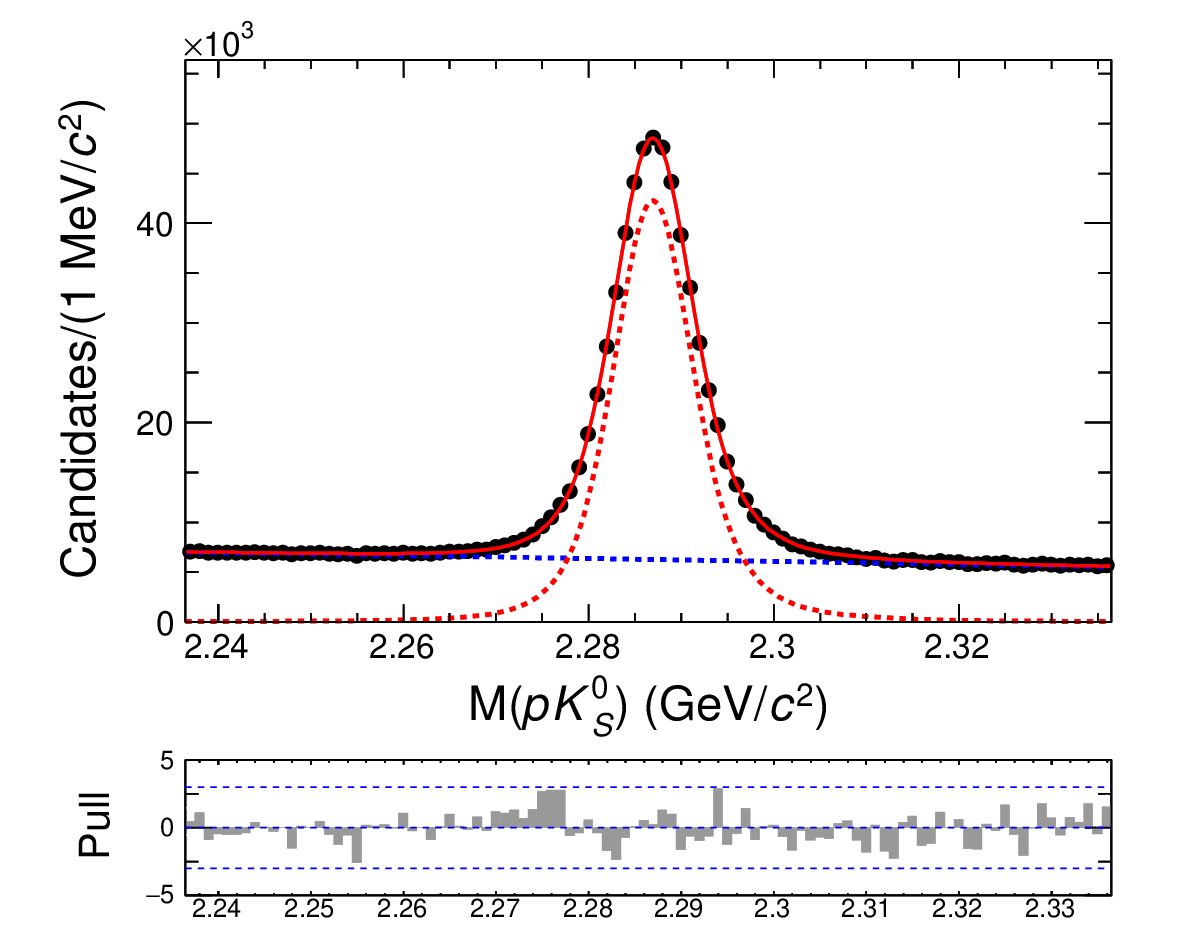}%
  \put(78,65){\large (c)}%
  \end{overpic}%
  \vskip-10pt
  \caption{\label{fig:Mfit_final}The distributions of invariant mass of $\Lcp$ candidates (points with error bars) and corresponding fit results (red curves) for (a) $\LcTopKsKs$, (b) $\LcTopKsEta$, and (c) $\LcTopKs$, respectively. The red (blue) dashed histograms show the signal (background).}
  \end{centering}
\end{figure*}

In fitting data, the mean is allowed a common shift ($\delta_\mu$) from the value found in MC, and the widths are those found in MC, multiplied by a common scaling factor ($k_{\sigma}$).
The background PDF is a first-order polynomial function for $\LcTopKsKs$ and a second-order polynomial function for $\LcTopKsEta$ and $\LcTopKs$.
The background parameters are floated to account for differences between the experimental data and MC simulated samples. 
The results are shown in Fig.~\ref{fig:Mfit_final}, along with the pulls $(N_{\rm data}-N_{\rm fit})/\sigma_{\rm data}$ where $\sigma_{\rm data}$ is the error on $N_{\rm data}$. 
The pull distributions demonstrate that the data are statistically consistent with the fitted shapes.
The signal and background yields are listed in Table~\ref{tab:MFit_unblind}.

For the $\LcTopKsKs$ mode, we obtain the difference in the log likelihoods obtained from fits performed with and without a signal PDF, ${\Delta\ln\mathcal{L}\!=\!524}$; as the number of degrees of freedom without a signal component is three less than that with a signal component (parameters $N_{\rm sig}$, $\delta_{\mu}$ and $k_{\sigma}$ are dropped), and this value of ${\Delta\ln\mathcal{L}}$ corresponds to a statistical significance greater than $10\sigma$. This measurement constitutes the first observation of this SCS $\Lcp$ decay. 

\begin{table}[!htbp]
\begin{centering}
\caption{\label{tab:MFit_unblind}The fitted yields of signal and background in the overall fit region (FR) and the signal region (SR) for the $\LcTopKsKs$, $\LcTopKsEta$, and $\LcTopKs$ modes. For the definition of these regions, see the text. The yields in signal region, $N_{\rm bkg}^{\rm SR}$ of $\Lcp\to p\KS(\KS,\eta)$ and $N_{\rm sig}^{\rm SR}$ of $\LcTopKs$, are used to measure the branching fractions.}
\setlength{\tabcolsep}{1.5mm}{
\begin{tabular}{cccc} \hline \hline 
 Yields	&  $\LcTopKsKs$ 		&   $\LcTopKsEta$		&  $\LcTopKs$  		\\ \hline   
 $N_{\rm sig}^{\rm FR}$ 	& $\,\,2442 \pm 103$ 		& $12877 \pm 317$    		& $515296 \pm 1129$ 	\\
 $N_{\rm bkg}^{\rm FR}$ 	& $41138 \pm 222$ 	  	& $75144 \pm 403$    		& $627427 \pm 1177$  	 \\
\hline 
 $N_{\rm sig}^{\rm SR}$ 	& $2391 \pm 101$ 		& $12641 \pm 311$    		& $500457 \pm 1096$  	\\
 $N_{\rm bkg}^{\rm SR}$  	& $8228 \pm 44$ 	  	& $32935 \pm 177$    		& $226055 \pm 424\,\,$ 	 \\ 
 \hline \hline 
\end{tabular}}
\end{centering}  
\end{table} 

\section{Branching fraction}
\begin{figure*}
  \begin{centering}%
  \begin{overpic}[width=0.333\textwidth]{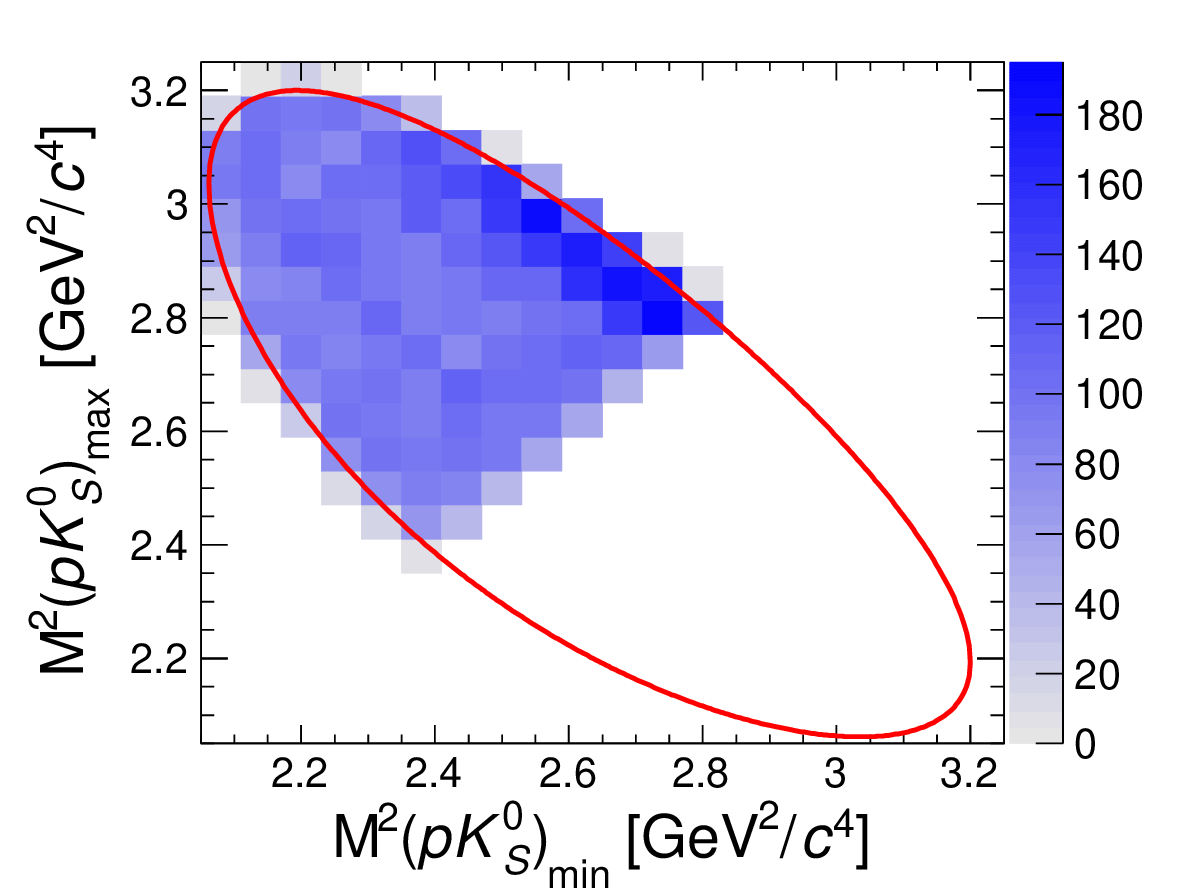}%
  \put(72,62){\large (a)}%
  \end{overpic}%
  \begin{overpic}[width=0.333\textwidth]{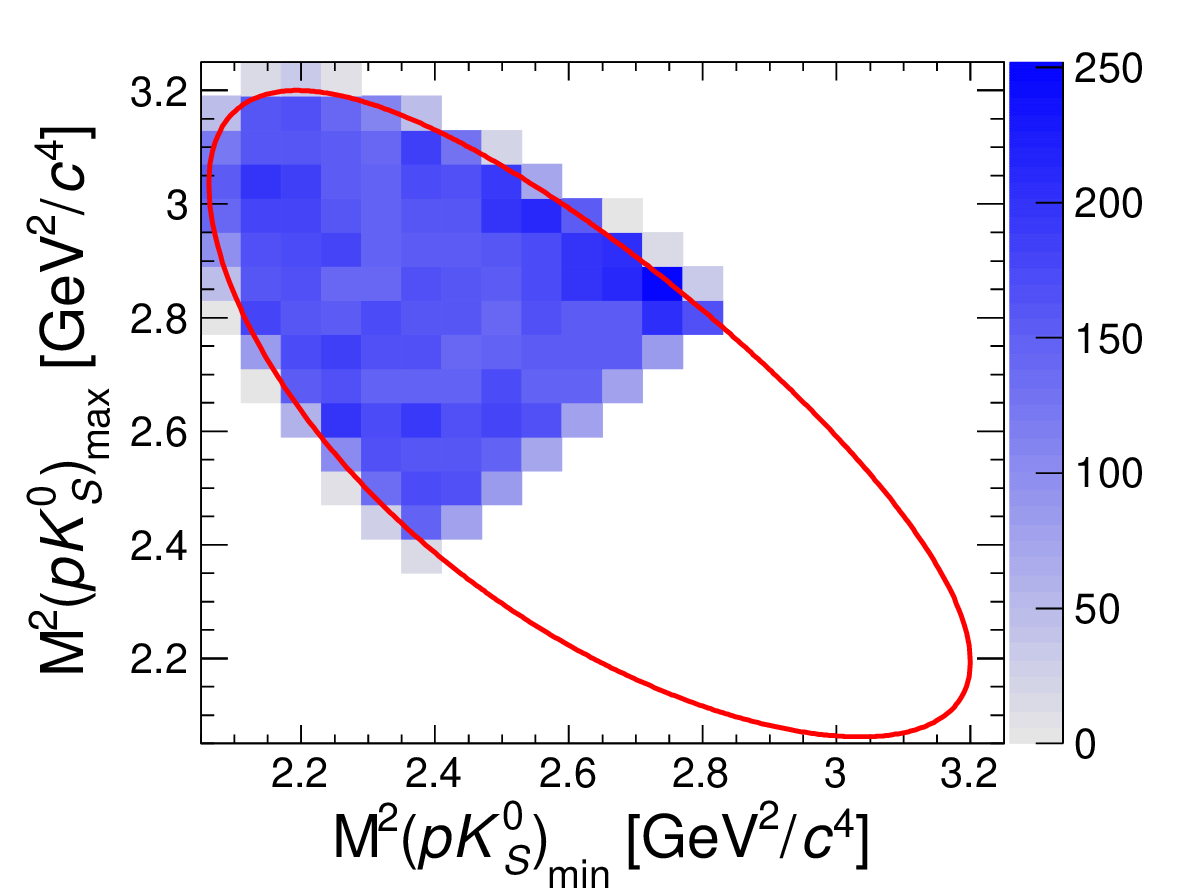}%
  \put(72,62){\large (b)}%
  \end{overpic}%
  \begin{overpic}[width=0.333\textwidth]{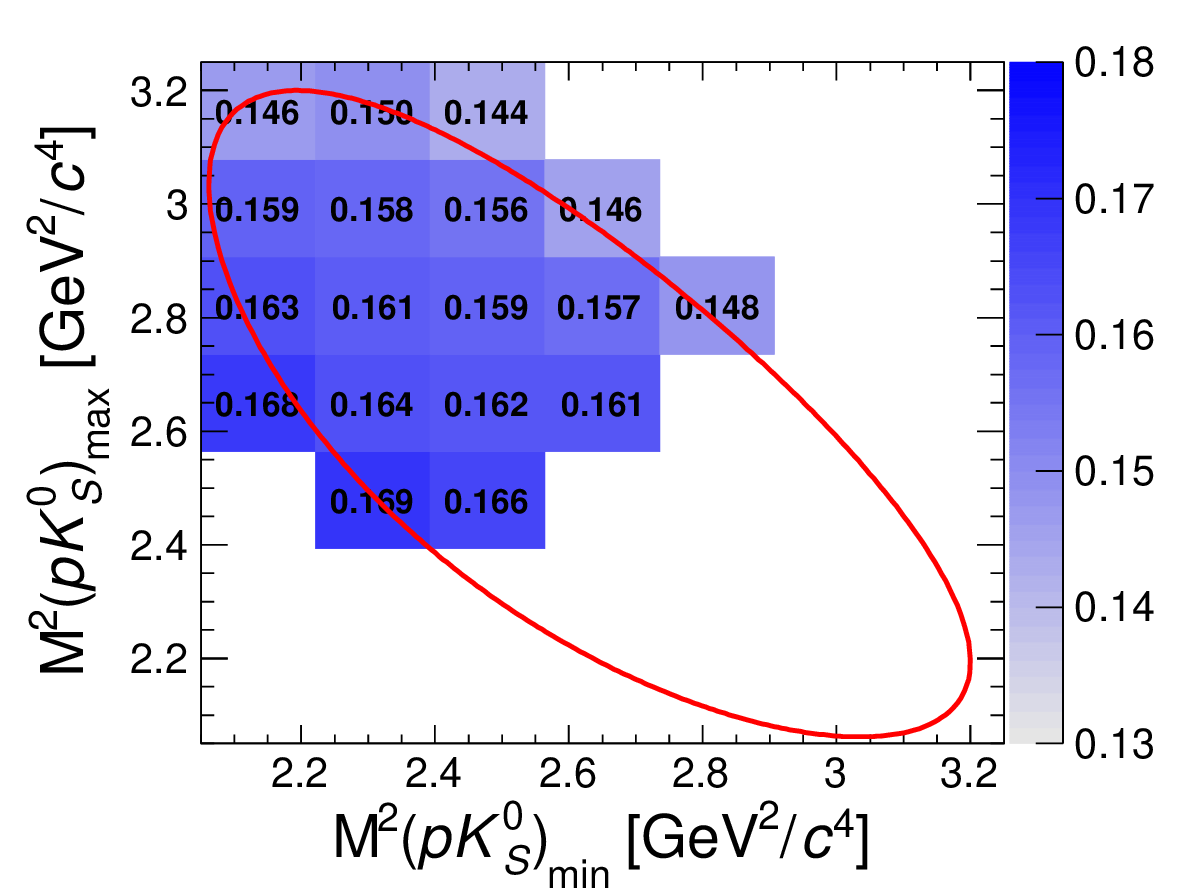}%
  \put(72,62){\large (c)}%
  \end{overpic}\\%
  \begin{overpic}[width=0.333\textwidth]{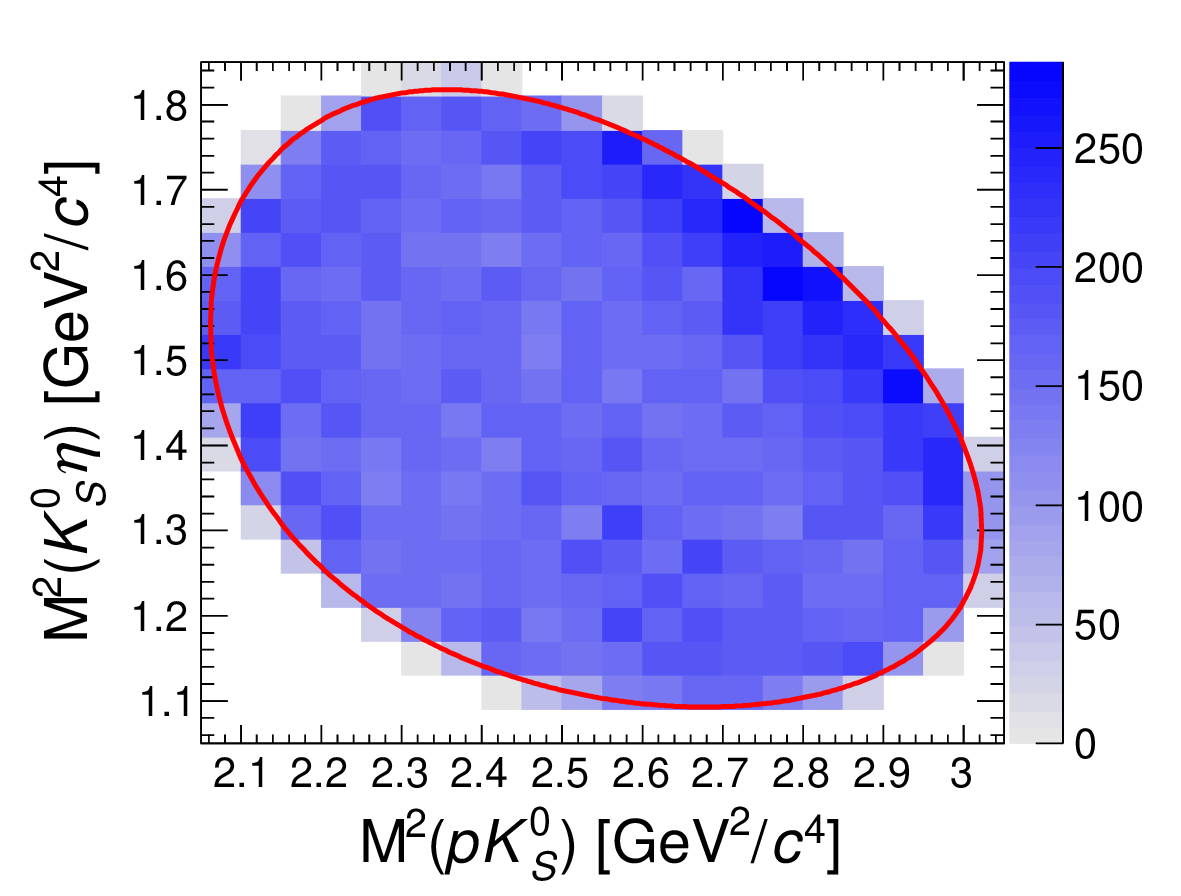}%
  \put(72,62){\large (d)}%
  \end{overpic}%
  \begin{overpic}[width=0.333\textwidth]{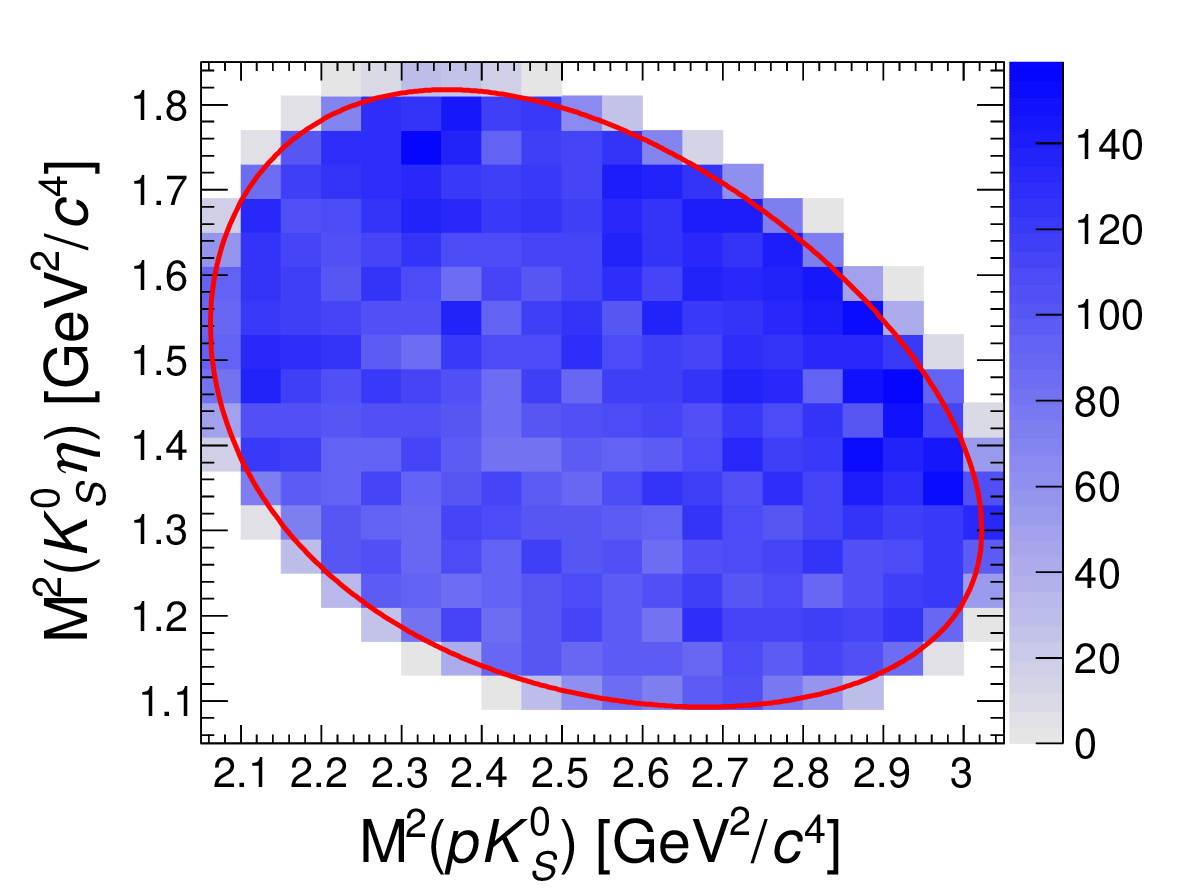}%
  \put(72,62){\large (e)}%
  \end{overpic}%
  \begin{overpic}[width=0.333\textwidth]{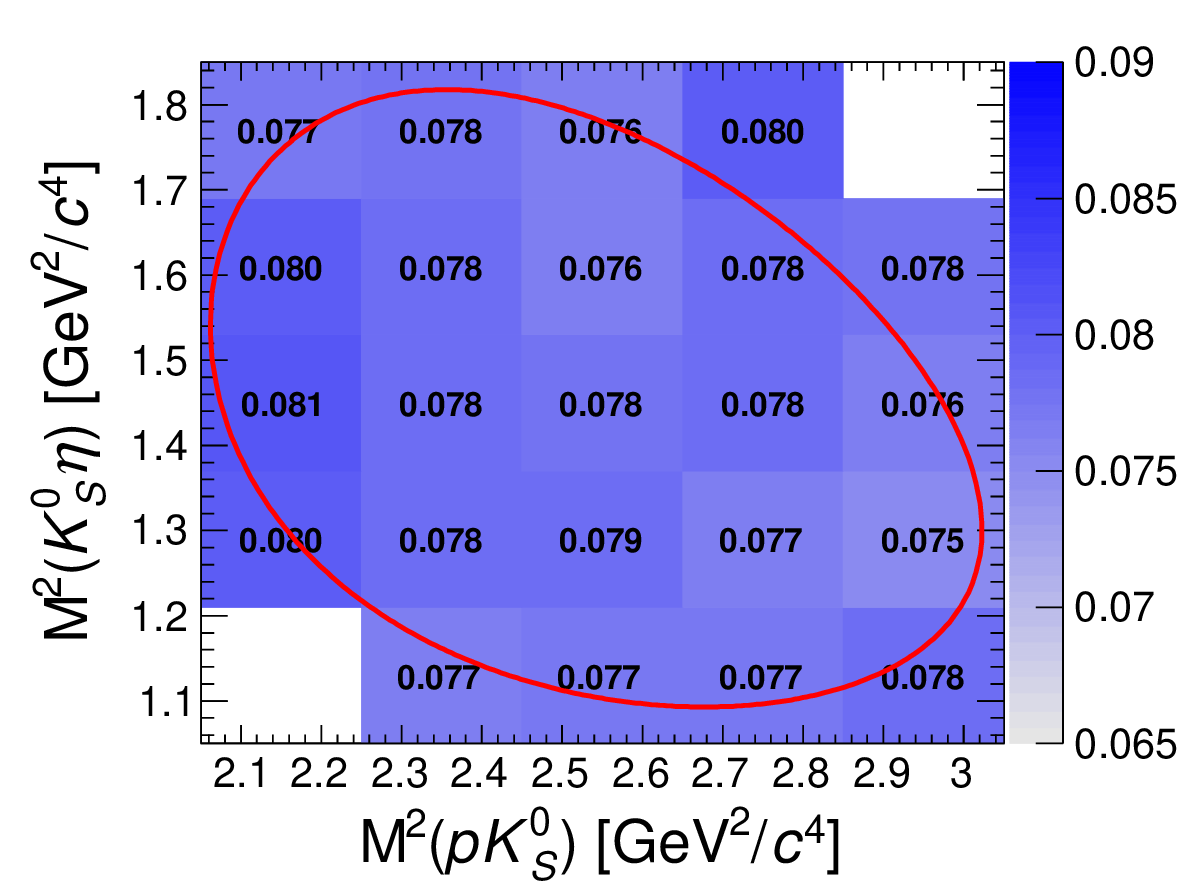}%
  \put(72,62){\large (f)}%
  \end{overpic}\\%
  \vskip-10pt
  \caption{\label{fig:effDP_Dlz}Plots (a,\,d) show the Dalitz plots in the $M(\Lcp)$ signal region, and (b,\,e) show the Dalitz plots in the $M(\Lcp)$ sideband region for $\LcTopKsKs$ (top) and $\LcTopKsEta$ (bottom). Plots (c,\,f) show the average signal efficiency in bins across the Dalitz plane. 
  The red curves show the edges of kinematic phase-space region of the decays.}
  \end{centering}
\end{figure*}

For the three-body decay modes, the Dalitz plots for candidates in the $M(\Lcp)$ signal region and sideband region are shown in Figs.~\ref{fig:effDP_Dlz}(a,\,b) for $\LcTopKsKs$ and Figs.~\ref{fig:effDP_Dlz}(d,\,e) for $\LcTopKsEta$. 
For $\LcTopKsKs$, Bose symmetry requires invariance under the exchange of the two $\KS$'s, 
hence the Dalitz plot for two $p\KS$ masses is symmetric. We plot $M^2(p\KS)_{\rm max}$ versus $M^2(p\KS)_{\rm min}$ in half of the Dalitz plot, as shown in Figs.~\ref{fig:effDP_Dlz}(a--c), and use it to measure the branching fraction.

For each mode, a large MC sample of signal events, generated uniformly across the decay phase space, is used to determine the reconstruction efficiency.
For $\LcTopKsKs$ and $\LcTopKsEta$, the efficiencies are calculated in bins across the phase space, based on truth-matched signal yield in the $M(\Lcp)$ signal region. 
The results are shown in Fig.~\ref{fig:effDP_Dlz}(c) for $\LcTopKsKs$ and Fig.~\ref{fig:effDP_Dlz}(f) for $\LcTopKsEta$.

In order to calculate the efficiency-corrected yield, properly taking into account the variations in efficiency and uncertainties in signal yield over the Dalitz plot, we make a bin-by-bin correction.
The Dalitz plots are divided uniformly into 7$\times$7 bins for $\LcTopKsKs$ and 5$\times$5 bins for $\LcTopKsEta$, as shown in Figs.~\ref{fig:effDP_Dlz}(c,\,f) respectively. The efficiency-corrected yields are 
\begin{eqnarray}
N_{\rm{corr}} & =  & \sum_i (N_i^{\rm tot}-N_{\rm bkg}^{\rm SR}f_i^{\rm{bkg}}) / \eff_i\,,  \label{eqn:correctedYeilds}
\end{eqnarray}
where $N_i^{\rm{tot}}$ is the raw yield in the $i^{\rm th}$ bin of the Dalitz plot in $M(\Lcp)$ signal region,
$N_{\rm bkg}^{\rm SR}$ is the fitted background yield as listed in Table~\ref{tab:MFit_unblind}, 
$f_i^{\rm{bkg}}$ is the fraction of background in the $i^{\rm th}$-bin, with $\sum_i{}f_i\!=\!1$. 
These fractions are obtained from the Dalitz plot distribution of events in the $M(\Lcp)$ sideband region, shown in Fig.~\ref{fig:effDP_Dlz}(b) for $\LcTopKsKs$ and Fig.~\ref{fig:effDP_Dlz}(e) for $\LcTopKsEta$. 
Using the generic MC sample, we find that the Dalitz plot in the chosen $M(\Lcp)$ sideband region is consistent with the generic background in the $M(\Lcp)$ signal region. 
The uncertainties on each 
variable in Eq.~(\ref{eqn:correctedYeilds}) have been considered and are propagated into the efficiency-corrected yields, $N_{\rm corr}$.
We obtain 
\begin{eqnarray}
N_{\rm{corr}}(\LcTopKsKs) & = & (1.55\pm0.08)\times10^4\,, \label{eqn:NcorResults1} \\
N_{\rm{corr}}(\LcTopKsEta) & = & (1.63\pm0.04)\times10^5\,. \label{eqn:NcorResults2}
\end{eqnarray}

The relative branching fractions of signal modes to reference mode are determined by Eqs.~(\ref{eqn:BRratioA},\,\ref{eqn:BRratioB}). 
\begin{small}
\begin{eqnarray}
\hskip-20pt
\frac{\mathcal{B}(\LcTopKsKs)}{\mathcal{B}(\LcTopKs)} 
& = &  \frac{N_{\rm{corr}}(\LcTopKsKs)}{\BR(\KS\to\pip\pim) N^{\rm SR}_{\rm{sig}}(\LcTopKs)/\eff_0}\,,  \label{eqn:BRratioA} \\
\hskip-20pt
\frac{\mathcal{B}(\LcTopKsEta)}{\mathcal{B}(\LcTopKs)} 
& = & \frac{N_{\rm{corr}}(\LcTopKsEta)}{\BR(\eta\to\gamma\gamma)N^{\rm SR}_{\rm{sig}}(\LcTopKs)/\eff_0}\,.  \label{eqn:BRratioB} 
\end{eqnarray}
\end{small}%
Here, $\eff_{0}\!=\!(33.09\pm 0.05)\%$ is the efficiency of the reference mode $\LcTopKs$ in the $M(\Lcp)$ signal region.
Inserting the efficiency-corrected yields in Eqs.~(\ref{eqn:NcorResults1},\,\ref{eqn:NcorResults2}), $N^{\rm SR}_{\rm{sig}}(\LcTopKs)$ in Table~\ref{tab:MFit_unblind},
and the world averages $\BR(\KS\to\pip\pim)\!=\!(69.20\pm 0.05)\%$ and $\BR(\eta\to\gamma\gamma)\!=\!(39.41\pm 0.20)\%$~\cite{bib:PDG2022}, 
we find 
\begin{eqnarray}
\frac{\mathcal{B}(\LcTopKsKs)}{\mathcal{B}(\LcTopKs)} & = &  (1.48\pm 0.08 )\times10^{-2}\,, \label{eqn:BRresult1_blind} \\
\frac{\mathcal{B}(\LcTopKsEta)}{\mathcal{B}(\LcTopKs)} & = &  (2.73\pm 0.06)\times10^{-1}\,,  \label{eqn:BRresult2_blind}
\end{eqnarray}
Combining with the world average branching fraction of reference mode $\BR(\LcTopKs)\!=\!(1.59\pm 0.08)\%$~\cite{bib:PDG2022}, we have the absolute branching fractions:
\begin{eqnarray}
\hskip-10pt
\mathcal{B}(\LcTopKsKs) & = & (2.35\pm 0.12 \pm 0.12)\times10^{-4}\,, 	\\
\hskip-10pt
\mathcal{B}(\LcTopKsEta) & = & (4.35\pm 0.10 \pm 0.22)\times10^{-3}\,, 
\end{eqnarray}
where the uncertainties are statistical and from the uncertainty on $\BR(\LcTopKs)$.

\begin{figure}
  \begin{centering}%
  \begin{overpic}[width=0.25\textwidth]{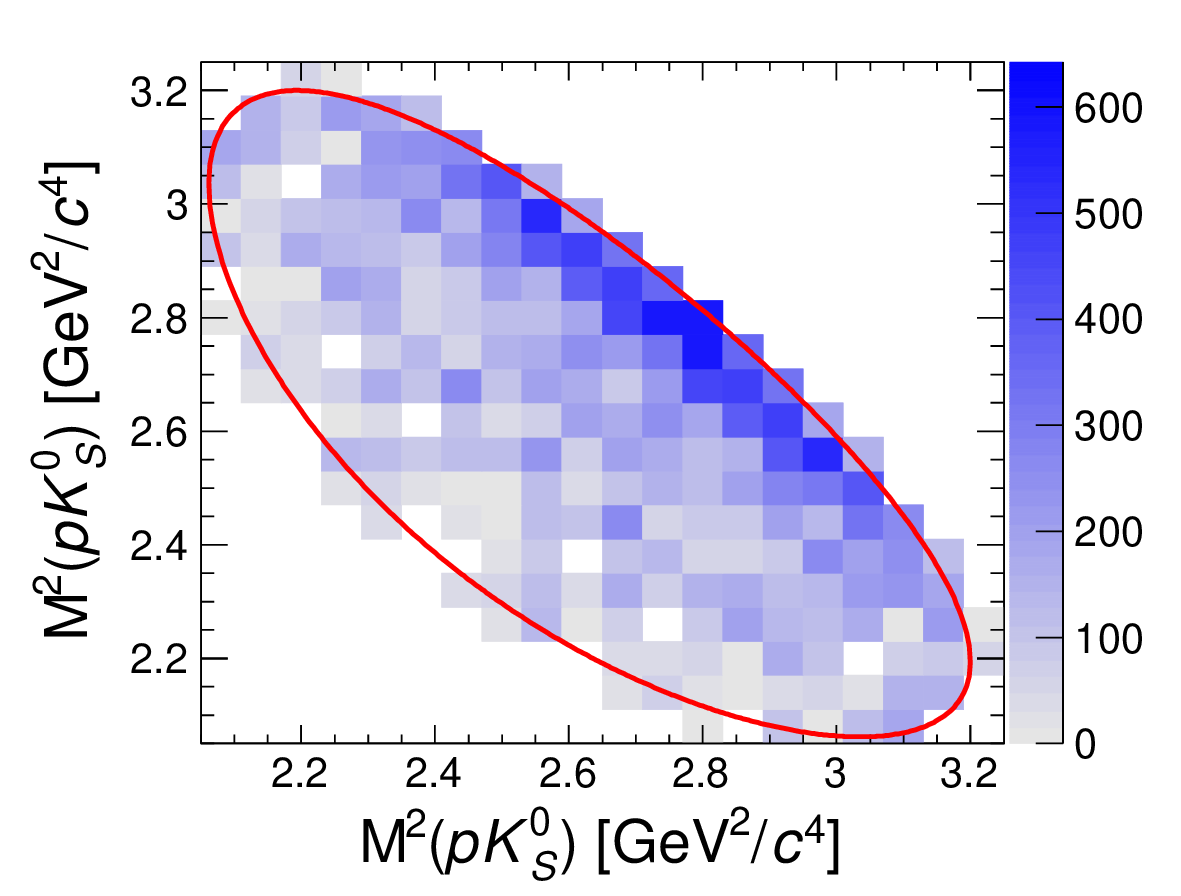}%
  \put(71,61){\large (a)}%
  \end{overpic}%
  \begin{overpic}[width=0.25\textwidth]{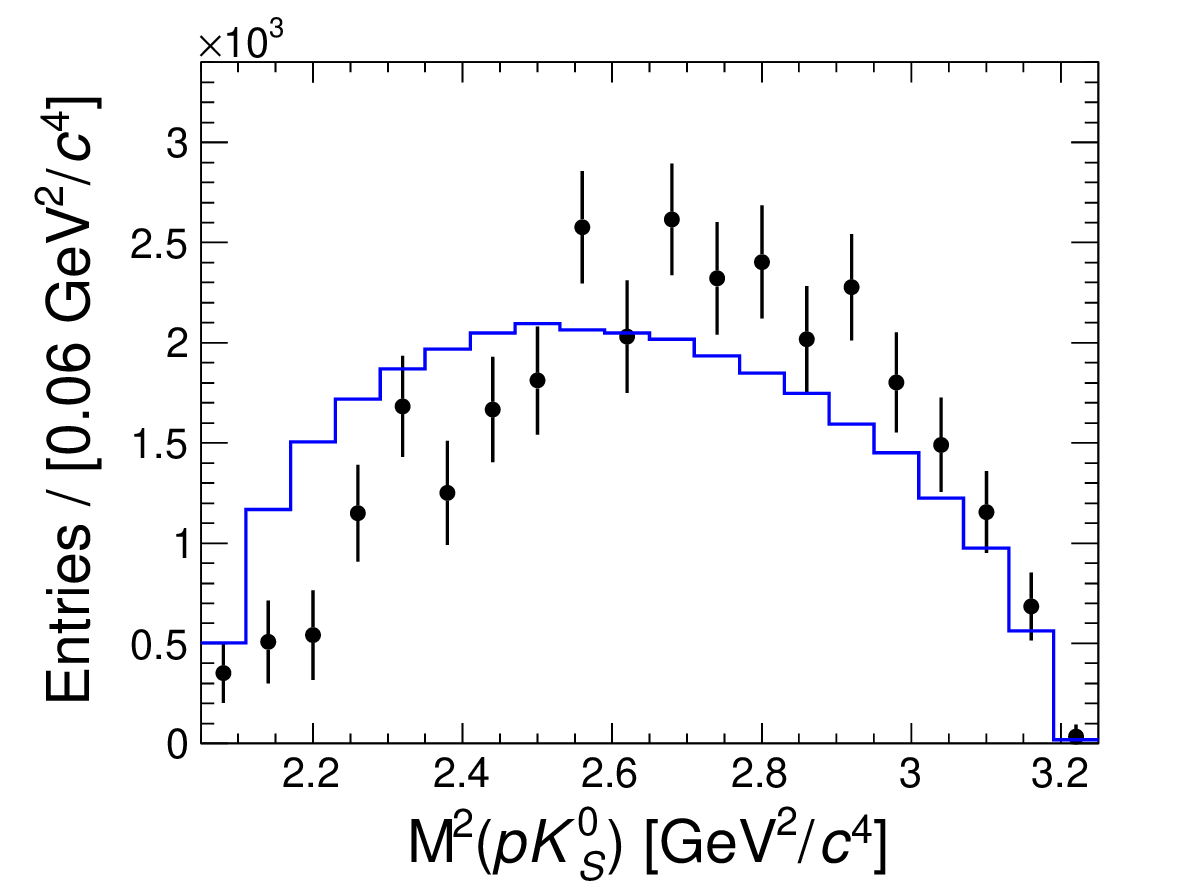}%
  \put(75,58){\large (b)}%
  \end{overpic}\\
  \begin{overpic}[width=0.25\textwidth]{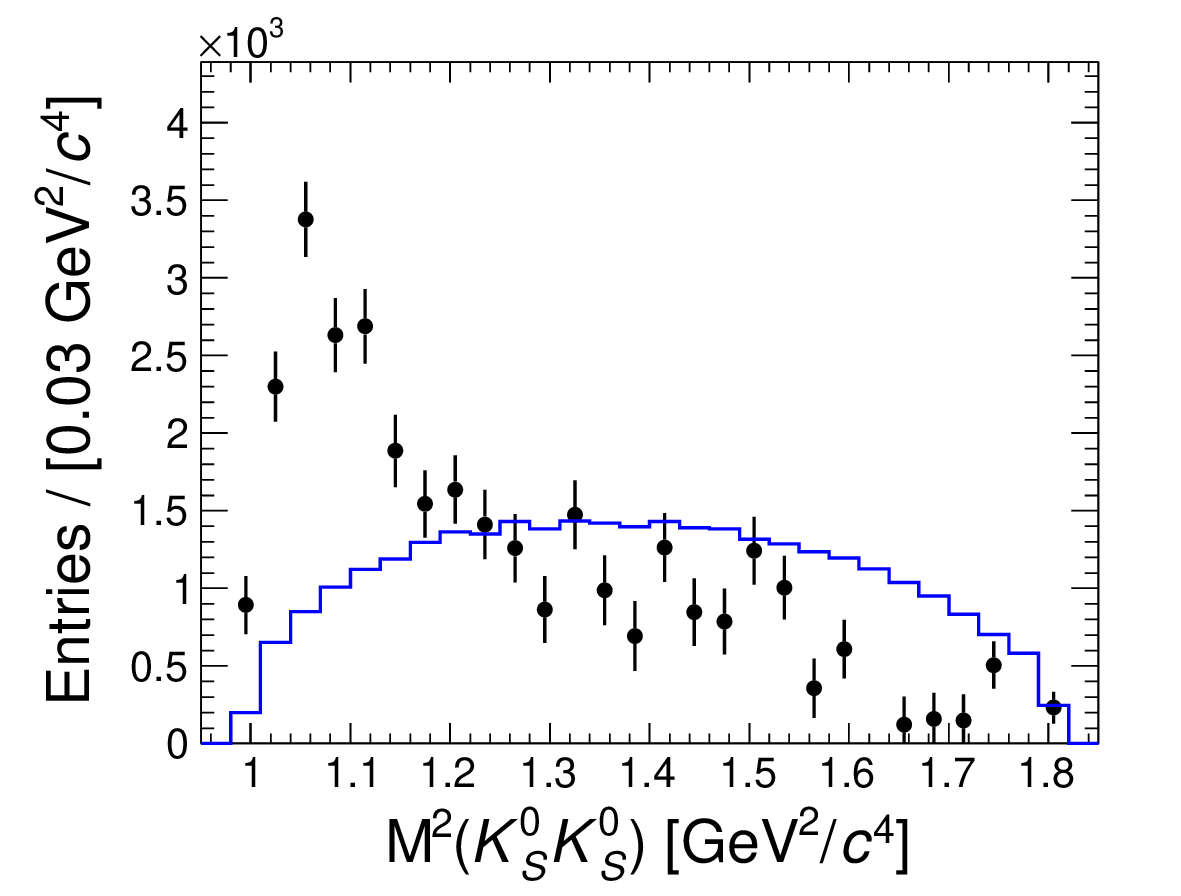}%
  \put(75,58){\large (c)}%
  \end{overpic}
  \vskip-10pt
  \caption{\label{fig:DP_LcTopKsKs}For $\LcTopKsKs$, the Dalitz plot after background subtraction and efficiency correction bin-by-bin and its projections superimposing with signal MC produced by phase space mode (blue histograms). This symmetric Dalitz plot and its projections show two entries per candidate, one for each possible $p\KS$ combination. A dominant structure near the $\KS\KS$ threshold, which we identify with $f_0(980)$ or $a_0(980)^0$, is clearly seen.}
  \end{centering}
\end{figure}

\begin{figure}
  \begin{centering}%
  \begin{overpic}[width=0.25\textwidth]{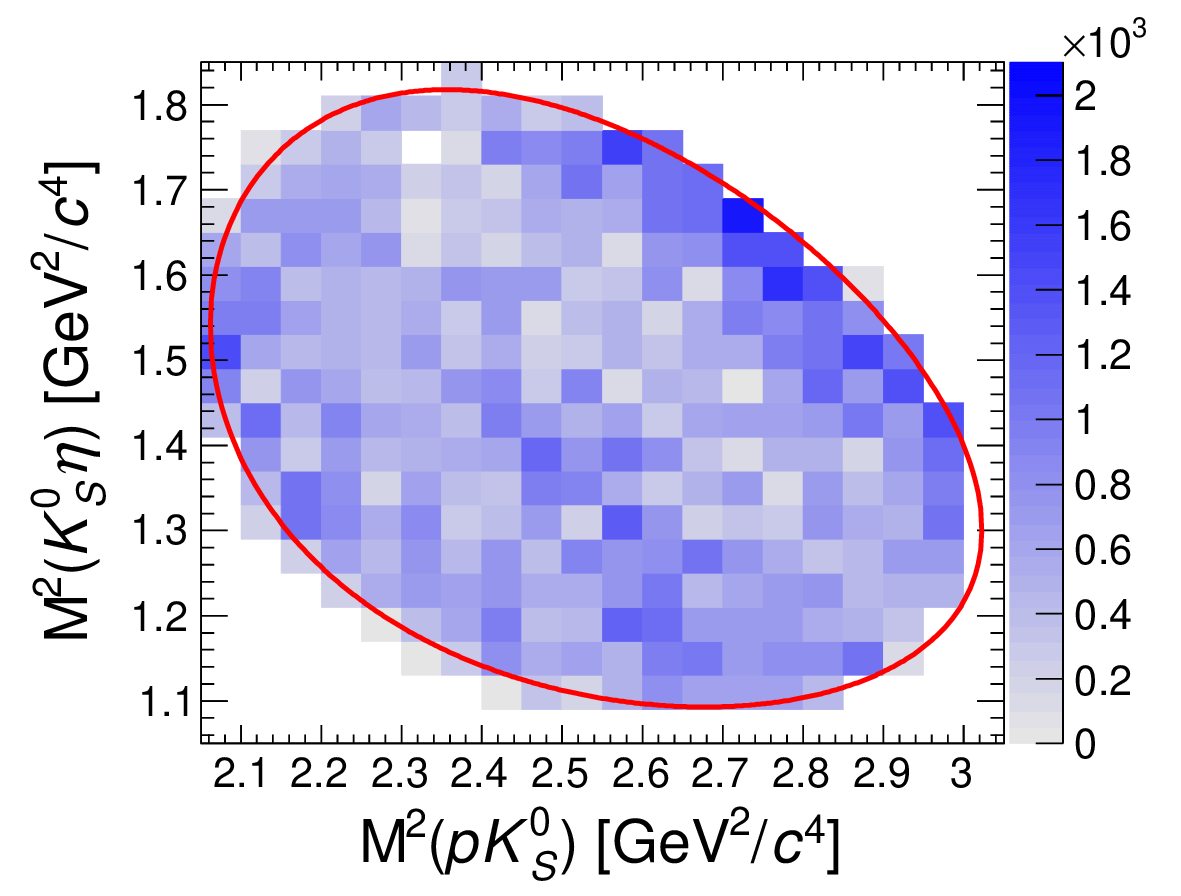}%
  \put(71,61){\large (a)}%
  \end{overpic}%
  \begin{overpic}[width=0.25\textwidth]{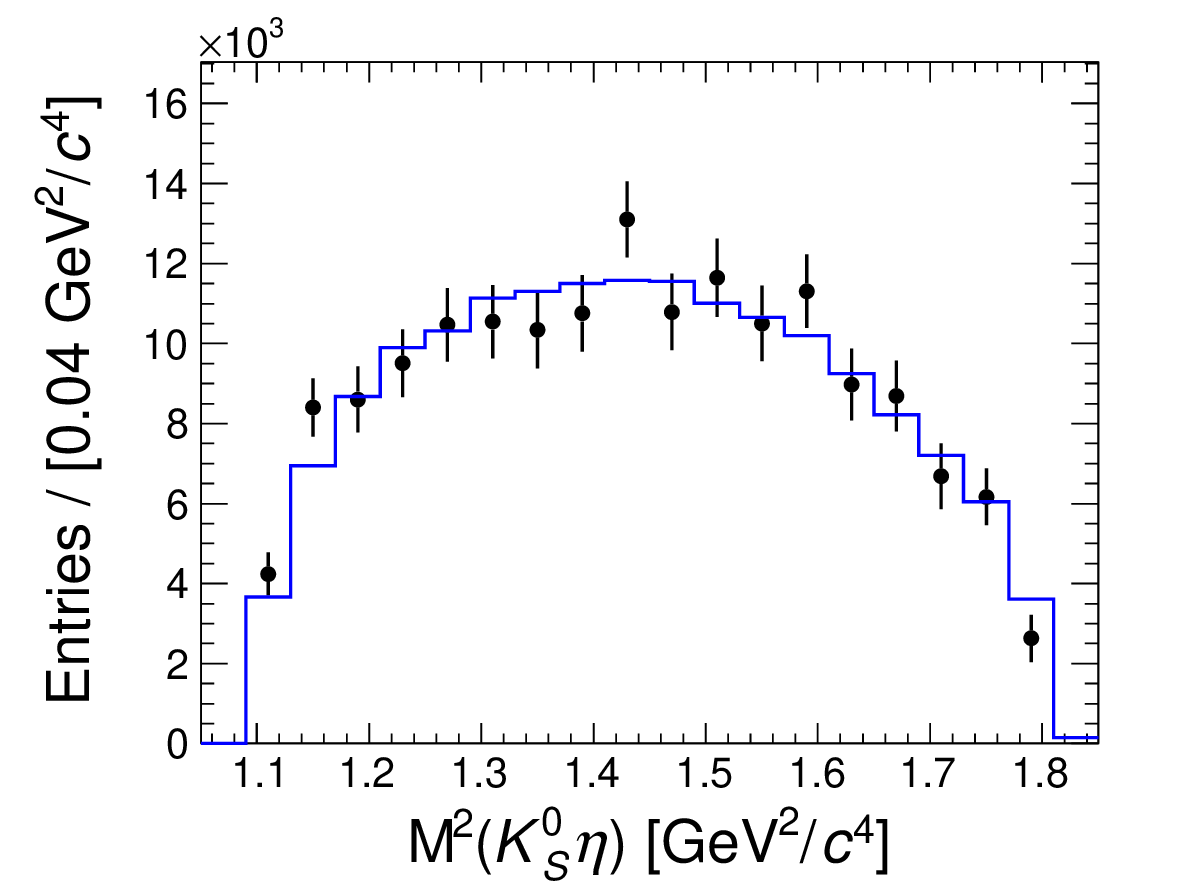}%
  \put(75,58){\large (b)}%
  \end{overpic}\\
  \begin{overpic}[width=0.25\textwidth]{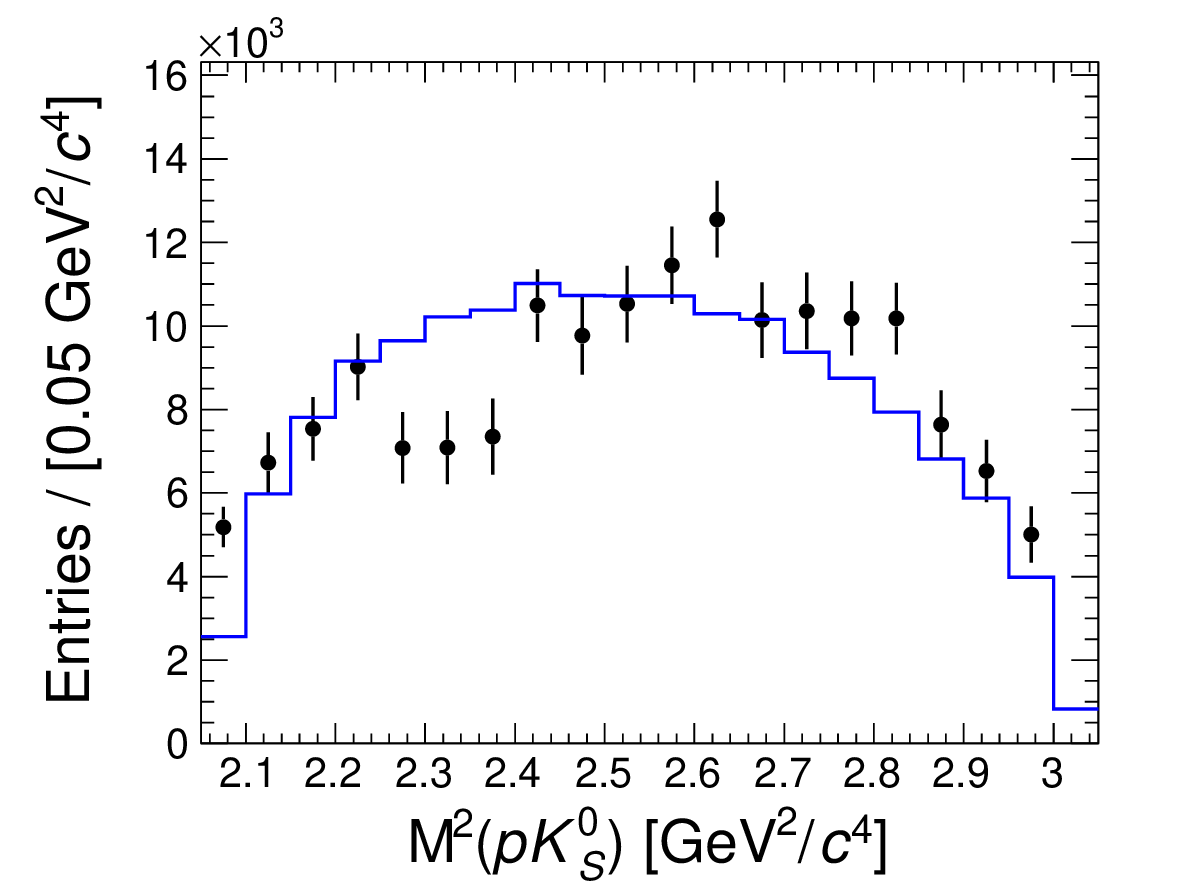}%
  \put(75,58){\large (c)}%
  \end{overpic}%
  \begin{overpic}[width=0.25\textwidth]{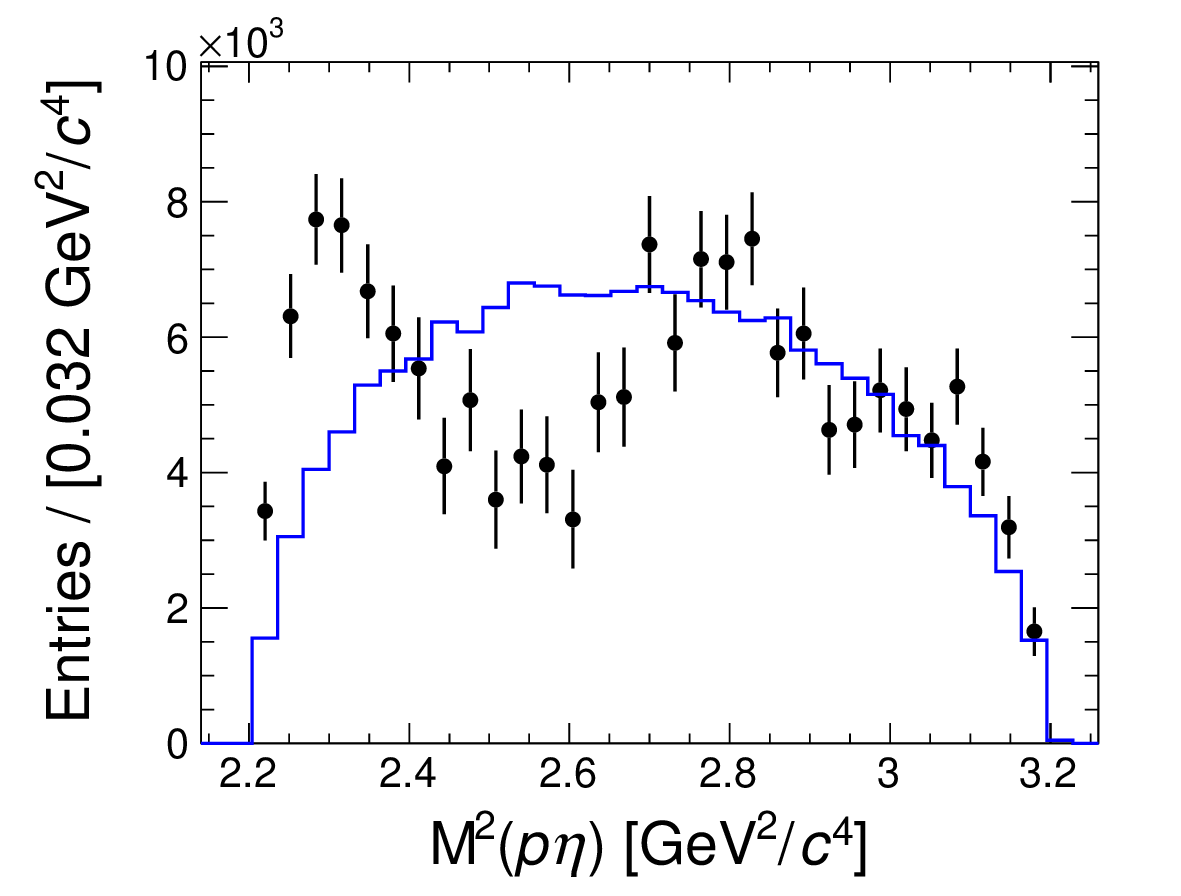}%
  \put(75,58){\large (d)}%
  \end{overpic}\\%
  \vskip-10pt
  \caption{\label{fig:DP_LcTopKsEta}For $\LcTopKsEta$, the Dalitz plot after background subtraction and efficiency correction bin-by-bin and its projections superimposing with signal MC produced by phase space mode (blue histograms). A significant structure of $N^{*}(1535)$ near the $p\eta$ threshold is found.}
  \end{centering}
\end{figure}

We examine the Dalitz plots for $\LcTopKsKs$ and $\LcTopKsEta$, after background subtraction and efficiency correction, for intermediate resonances. 
In $\LcTopKsKs$, clear evidence for $f_0(980)$ or $a_0(980)^0$ (labeled as $S_0(980)$) near the $\KS\KS$ threshold is seen, as shown in Fig.~\ref{fig:DP_LcTopKsKs}.
In $\LcTopKsEta$, a significant enhancement  consistent with $N^*(1535)$ is found near the $p\eta$ threshold, as shown in Fig.~\ref{fig:DP_LcTopKsEta}. 
In the future, amplitude analyses of these decays  can be expected to improve our understanding of the nature of $S_{0}(980)$ and $N^{*}(1535)$.

\section{Systematic uncertainty}
In measuring the ratio of branching fractions, many systematic uncertainties cancel, as they affect both the signal and reference modes. 
The remaining systematic uncertainties are summarized in Table~\ref{tab:sysBR} and introduced in detail below.  
\begin{table}[!htbp]
\begin{centering}
\caption{\label{tab:sysBR}Relative systematic uncertainties of the branching fractions of $\LcTopKsKs$ and $\LcTopKsEta$, and the uncertainty from the branching fraction of the reference mode.}
\setlength{\tabcolsep}{0.5mm}{
\begin{tabular}{lcc} \hline \hline 
 sources 	& $\mathcal{B}_{\LcTopKsKs}$  & $\mathcal{B}_{\LcTopKsEta}$ \\ \hline  
$\KS$ reconstruction	&	1.4\%	&	0.4\%	\\
proton PID efficiency   & 	0.9\%	&   0.5\%	\\
$\eta$ reconstruction	&	--	    &	4.0\%	\\				
$M(\Lcp)$ fit procedure      
                        &   1.9\%   &   2.3\%   \\
efficiency-correction procedure 
                        &   0.8\%   &   0.4\%   \\
non-$\KS$ peaking background 
                        &   0.8\%	&	--	    \\
$\delta\BR/\BR(\KS\to\pip\pim,\,\eta\to\gamma\gamma)$ 	
                        &  	0.1\%	&	0.5\%	\\ \hline
total syst.	uncertainty	&	2.8\%	&	4.7\%		\\ \hline 
$\delta\BR/\mathcal{B}(\LcTopKs)$	
				        & 	5.0\%	& 	5.0\%	\\ \hline 
\hline 
\end{tabular}}
\end{centering}  
\end{table} 

The systematic uncertainty associated with the $\KS$ reconstruction is considered as follows.  
A table of $\KS$ efficiency ratios of data to MC in eight bins of the $\KS$ momentum distribution, $R{}_{\eff}^{\KS}$, 
is determined based on a control sample ${D^{*\pm}\to (D^0\to\KS\piz)\pi^{\pm}}$.
The unfolded momentum distribution in data of $\KS$ from signal is obtained using the $\sPlot$ technique~\cite{Pivk:2004ty}. 
From one $R{}_{\eff}^{\KS}$ table, 
we can determine the average ratios:
(1)\,for $\Lcp\to{}pK^{0}_{S,{\rm fast}}K^{0}_{S,{\rm slow}}$ where the subscript `fast' (`slow') indicates the faster (slower) of two $\KS$'s in the final state, $\overline{R}{}_{\eff}^{\KS} \!=\! \sum_i^8\sum_j^8 N_{ij}  (R{}_{\eff,i}^{K^0_{S,{\rm fast}}} R{}_{\eff,j}^{K^{0}_{S,{\rm slow}}}) / \sum_i^8\sum_j^8 N_{ij}$ calculated on the two-dimensional $(p_{\scriptscriptstyle{K^0_{S,{\rm fast}}}},\,p_{\scriptscriptstyle{K^0_{S,{\rm slow}}}})$ distribution due to the correlations between the momenta of two $\KS$'s. 
Here $N_{ij}$ and $(R{}_{\eff,i}^{K^0_{S,{\rm fast}}}R{}_{\eff,j}^{K^0_{S,{\rm slow}}})$ are the yield and the averaged $R{}_{\eff}^{\KS}$, respectively, in the bin of $i^{\rm th}$ raw and $j^{\rm th}$ column of such two-dimensional momenta distribution;
(2)\,for $\LcTopKs(\eta)$, ${\overline{R}{}_{\eff}^{\KS} \!=\! \sum_i^8 N_i R{}_{\eff,i}^{\KS}/\sum_i^8 N_i}$ calculated on the one-dimensional $p_{\scriptscriptstyle\KS}$ distribution. 
Here $N_i$ and $R{}_{\eff,i}^{\KS}$ are the yield and the averaged $R{}_{\eff}^{\KS}$, respectively, in the $i^{\rm th}$ bin of such one-dimensional distribution.
We build 10000 $R{}_{\eff}^{\KS}$ tables by randomly fluctuating $R{}_{\eff,i}^{\KS}$ in each bin according to its uncertainty and calculate $\overline{R}{}_{\eff}^{\KS}$ for each. 
We take the mean and root-mean-square (RMS) values from the distribution of $\overline{R}{}_{\eff,\,{\rm sig.}}^{\KS}/\overline{R}{}_{\eff,\,{\rm ref.}}^{\KS}-1$, where the subscripts `sig.' and `ref.' refer to the signal and reference modes, respectively, and add in quadrature as the estimate of the systematic uncertainty. 

Since the protons in the signal and reference modes have different kinematic distributions, the systematic effects due to PID do not cancel completely. 
The data/MC ratio of proton PID efficiency depends on the proton momentum and polar angle: $R{}_{\eff}^{p}(p,\cos\theta)$. 
Such a $R{}_{\eff}^{p}$ map is determined based on an inclusive sample of $\Lambda\to{}p\pim$. 
Following steps similar to those used above for $\KS$ efficiency, we obtain the unfolded ($p,\,\cos\theta$) two-dimensional distribution for protons using the $\sPlot$ technique~\cite{Pivk:2004ty}, and plot the $\overline{R}{}_{\eff,\,{\rm sig.}}^{p}/\overline{R}{}_{\eff,\,{\rm ref.}}^{p}\!-\!1$ values based on 10000 maps of $R_{\eff}^{p}(p,\,\cos\theta)$. 
The systematic uncertainty due to PID is obtained by adding in quadrature the mean and RMS values of the $\overline{R}{}_{\eff,\,{\rm sig.}}^{p}/\overline{R}{}_{\eff,\,{\rm ref.}}^{p}-1$ distribution.

The uncertainty due to $\eta\to\gamma\gamma$ reconstruction is estimated to be 4\%, considering 2\% per photon according to a study of radiative Bhabha events.

The systematic uncertainties from the $M(\Lcp)$ fits for $\LcTopKsKs$ and $\LcTopKsEta$ channels are evaluated to be 1.8\% and 2.3\%, respectively, after considering two sources below. 
(a) The uncertainty due to fixing the signal parameters in the fits is estimated by randomly varying them via a multiple-dimensional Gaussian function (including these parameters' uncertainties and their correlation matrix from the $M(\Lcp)$ fit of truth-matched signals).
We produce 1000 sets of such signal parameters and repeat the $M(\Lcp)$ fits. 
We take the ratio of RMS to mean value of the distribution of fitted yield as the relative systematic uncertainty: 0.2\% for $\LcTopKsKs$, 0.4\% for $\LcTopKsEta$, and 0.2\% for $\LcTopKs$. 
(b) To evaluate the potential fit bias, we perform a bias check for the fitted signal yield based on 1000 sets of MC samples, of which the signals are randomly sampled from a large signal MC sample and the backgrounds from the generic $B\overline{B}$ and continuum MC samples. Their sampled yields are equal to the fitted yields in Table~\ref{tab:MFit_unblind}. 
We perform $M(\Lcp)$ fits for these samples. The fitted signal yields are plotted and fitted with a Gaussian function. 
The shifts of the fitted mean values of the Gaussian functions from the corresponding input values are assigned as systematic uncertainties: 1.9\% for $\LcTopKsKs$, 2.3\% for $\LcTopKsEta$, and 0.1\% for $\LcTopKs$.
The uncertainties for signal modes and reference mode are added in quadrature, as listed in Table~\ref{tab:sysBR}.

The systematic effects from the efficiency corrections for the $\LcTopKsKs$ and $\LcTopKsEta$ channels are evaluated to be 
0.8\% and 0.4\%, 
respectively, which are obtained by taking the quadratic sum of the following sources:
(a) Varying bin size:
the 7$\times$7 bins are changed to 6$\times$6 and 8$\times$8 bins for $\LcTopKsKs$
and the 5$\times$5 bins are changed to 4$\times$4 and 6$\times$6 bins for $\LcTopKsEta$. 
The changes of efficiency-corrected yields, $0.2\%$ for $\LcTopKsKs$ and $0.1\%$ for $\LcTopKsEta$, 
are assigned as the systematic uncertainties.
(b) To estimate the uncertainties due to the background Dalitz plot, we shift the $M(\Lcp)$ sideband region by $\pm5$ MeV, and repeat the efficiency correction. 
The resulting changes of efficiency-corrected yields, 0.1\% for both channels, 
are assigned as systematic uncertainty. 
(c) The signal efficiency effects due to the additional requirements in the signal mode with respect to the reference mode, such as $p(\eta)$, $\chi^2_m(\eta)$, and $L/\sigma_L(\KS)$, are neglected, as the signal distributions unfolded from data using the $\sPlot$ technique~\cite{Pivk:2004ty} and truth-matched signal distributions from MC are consistent. 
(d) Systematic effects from the $\chi_{\rm vtx}^2$ requirement are considered, since the signal and reference modes have different $\chi_{\rm vtx}^2$ distributions. 
We change the requirement to $\chi_{\rm vtx}^2\!<\!21$ and repeat our measurement. The resulting changes to the nominal results, 0.6\% and 0.3\%, are small as expected and assigned as the corresponding systematic uncertainties.
(e) The uncertainty due to the $\piz$ veto for $\eta$ candidates in $\LcTopKsEta$ is estimated by enlarging the veto region from $\pm12.5$ MeV/$c^2$ to be $\pm15$ MeV/$c^2$. 
The resulting change on the branching fraction is 0.2\%, and is assigned as a systematic uncertainty. 
(f) The uncertainty due to possible data/MC differences in $M(\Lcp)$ resolution is estimated as follows. 
Defining $R$ as the ratio of the signal yield in the $M(\Lcp)$ signal region to that in the fit region, we calculate $r\!=\!R_{\rm data}/R_{\rm MC}$ for the signal and reference modes. The fractional difference in $r$ between signal and reference modes and the uncertainty thereon are summed in quadrature and taken as the systematic uncertainty, which we find to be 0.5\% for $\BR(\LcTopKsKs)/\BR(\LcTopKs)$ and 0.1\% for $\BR(\LcTopKsEta)/\BR(\LcTopKs)$.
(g) The uncertainty due to limited MC statistics for the efficiency value is 0.1\%.

The uncertainty due to the non-$\KS$ peaking background is estimated 
based on the generic MC sample aforementioned. 
As the rate of this background may depend on intermediate processes, we double its size, and take the resulting ratio with the signal yield, 0.8\%, as the associated systematic uncertainty.
The uncertainties on $\BR(\KS\to\pip\pim)\!=\!{(69.20\pm 0.05)\%}$ ($\delta\BR/\BR=0.1\%$) and $\BR(\eta\to\gamma\gamma)\!=\!{(39.41\pm 0.20)\%}$ ($\delta\BR/\BR=0.5\%$) are also considered.
All uncertainties above are added in quadrature to give an overall systematic uncertainty, as listed in Table~\ref{tab:sysBR}. Additionally, the uncertainty from the world average branching fraction of the reference mode ($5.0\%$) is considered.

\section{Summary}
In summary, based on the entire dataset with integrated luminosity $980~\invfb$ collected by the Belle detector at the KEKB energy-asymmetric $e^+e^-$ collider,
we present the first observation of the SCS decay $\LcTopKsKs$ with a statistical significance of ${>\!10\sigma}$ 
and measure the branching fractions of $\LcTopKsKs$ and $\LcTopKsEta$ relative to $\LcTopKs$: 
\begin{eqnarray}
\hskip-20pt
\frac{\mathcal{B}(\LcTopKsKs)}{\mathcal{B}(\LcTopKs)} & = & (1.48\pm 0.08\pm 0.04)\times10^{-2}\,, \\
\hskip-20pt
\frac{\mathcal{B}(\LcTopKsEta)}{\mathcal{B}(\LcTopKs)} & = & (2.73\pm 0.06\pm 0.13)\times10^{-1}\,,
\end{eqnarray}
where the uncertainties are statistical and systematic, respectively, 
Using the world average $\BR(\LcTopKs)=(1.59\pm 0.08)\%$~\cite{bib:PDG2022}, we obtain the absolute branching fractions 
\begin{eqnarray}
\mathcal{B}(\LcTopKsKs)  =  \hskip120pt   \nonumber \\
                   \hskip40pt (2.35\pm 0.12 \pm 0.07\pm 0.12) \times10^{-4}\,, \\
\mathcal{B}(\LcTopKsEta) = \hskip129pt   \nonumber \\
                    \hskip40pt (4.35\pm 0.10 \pm 0.20\pm 0.22)\times10^{-3}\,, 
\end{eqnarray}
where the first uncertainties are statistical, the second systematic, and the third from the uncertainty on $\mathcal{B}(\LcTopKs)$.
The first of these branching fractions is measured for the first time and found to be much smaller than the theoretical prediction of $(1.9\pm0.4)\!\times\!10^{-3}$~\cite{Cen:2019ims}. 
The latter is consistent with the world average, $(4.15\pm0.90)\!\times\!10^{-3}$~\cite{bib:PDG2022}, with a threefold improvement in precision.

We reconstruct the Dalitz plots for $\LcTopKsKs$ and $\LcTopKsEta$, with background subtractions and efficiency corrections.
We note two clear structures that are consistent with $f_0(980)\to \KS\KS$ or $a_0(980)\to \KS\KS$  and $N^*(1535)\to p\eta$, raising the expectation that the nature of these intermediate resonances will be probed in the future with amplitude analyses on the larger data sets anticipated from BESIII~\cite{BESIII:2020nme} and Belle II~\cite{Belle-II:2018jsg}. 

\section*{Acknowledgments}
This work, based on data collected using the Belle detector, which was
operated until June 2010, was supported by 
the Ministry of Education, Culture, Sports, Science, and
Technology (MEXT) of Japan, the Japan Society for the 
Promotion of Science (JSPS), and the Tau-Lepton Physics 
Research Center of Nagoya University; 
the Australian Research Council including grants
DP180102629, 
DP170102389, 
DP170102204, 
DE220100462, 
DP150103061, 
FT130100303; 
Austrian Federal Ministry of Education, Science and Research (FWF) and
FWF Austrian Science Fund No.~P~31361-N36;
the National Natural Science Foundation of China under Contracts
No.~11675166,  
No.~11705209;  
No.~11975076;  
No.~12135005;  
No.~12175041;  
No.~12161141008; 
Key Research Program of Frontier Sciences, Chinese Academy of Sciences (CAS), Grant No.~QYZDJ-SSW-SLH011; 
the Ministry of Education, Youth and Sports of the Czech
Republic under Contract No.~LTT17020;
the Czech Science Foundation Grant No. 22-18469S;
Horizon 2020 ERC Advanced Grant No.~884719 and ERC Starting Grant No.~947006 ``InterLeptons'' (European Union);
the Carl Zeiss Foundation, the Deutsche Forschungsgemeinschaft, the
Excellence Cluster Universe, and the VolkswagenStiftung;
the Department of Atomic Energy (Project Identification No. RTI 4002) and the Department of Science and Technology of India; 
the Istituto Nazionale di Fisica Nucleare of Italy; 
National Research Foundation (NRF) of Korea Grant
Nos.~2016R1\-D1A1B\-02012900, 2018R1\-A2B\-3003643,
2018R1\-A6A1A\-06024970, RS\-2022\-00197659,
2019R1\-I1A3A\-01058933, 2021R1\-A6A1A\-03043957,
2021R1\-F1A\-1060423, 2021R1\-F1A\-1064008, 2022R1\-A2C\-1003993;
Radiation Science Research Institute, Foreign Large-size Research Facility Application Supporting project, the Global Science Experimental Data Hub Center of the Korea Institute of Science and Technology Information and KREONET/GLORIAD;
the Polish Ministry of Science and Higher Education and 
the National Science Center;
the Ministry of Science and Higher Education of the Russian Federation, Agreement 14.W03.31.0026, 
and the HSE University Basic Research Program, Moscow; 
University of Tabuk research grants
S-1440-0321, S-0256-1438, and S-0280-1439 (Saudi Arabia);
the Slovenian Research Agency Grant Nos. J1-9124 and P1-0135;
Ikerbasque, Basque Foundation for Science, Spain;
the Swiss National Science Foundation; 
the Ministry of Education and the Ministry of Science and Technology of Taiwan;
and the United States Department of Energy and the National Science Foundation.
These acknowledgements are not to be interpreted as an endorsement of any
statement made by any of our institutes, funding agencies, governments, or
their representatives.
We thank the KEKB group for the excellent operation of the
accelerator; the KEK cryogenics group for the efficient
operation of the solenoid; and the KEK computer group and the Pacific Northwest National
Laboratory (PNNL) Environmental Molecular Sciences Laboratory (EMSL)
computing group for strong computing support; and the National
Institute of Informatics, and Science Information NETwork 6 (SINET6) for
valuable network support.
We thank Li-Sheng Geng and Ju-Jun Xie for helpful discussions on the $N^{*}(1535)$.

\bibliography{references.bib}
\end{document}